\documentclass[aps, prl, reprint, superscriptaddress, showpacs,preprintnumbers,amsmath,amssymb,twocolumn]{revtex4-2}
\usepackage{amssymb}
\usepackage{amsmath}
\usepackage{bm}
\usepackage[dvips]{graphicx}
\usepackage{dcolumn}
\usepackage{longtable}
\usepackage{color}
\usepackage[version = 4]{mhchem}

\begin{document}
\title{Change of superconducting character in \ce{UTe2} induced by magnetic field}
\author{K.~Kinjo}
\email{kinjo.katsuki.63v@st.kyoto-u.ac.jp}
\author{H.~Fujibayashi}
\author{S.~Kitagawa}
\author{K.~Ishida}
\email{kishida@scphys.kyoto-u.ac.jp}
\affiliation{Department of Physics, Graduate School of Science, Kyoto University, Kyoto 606-8502, Japan}
\author{Y.~Tokunaga}
\author{H.~Sakai}
\author{S.~Kambe}
\affiliation{Advanced Science Research Center, Japan Atomic Energy Agency, Tokai, Ibaraki 319-1195, Japan}
\author{A.~Nakamura}
\author{Y.~Shimizu}
\author{Y.~Homma}
\author{D.~X.~Li}
\author{F.~Honda}
\affiliation{Institute for Materials Research, Tohoku University, Oarai, Ibaraki 311-1313, Japan}
\affiliation{Central Institute of Radioisotope Science and Safety, Kyushu University, Fukuoka 819-0395, Japan}
\author{D.~Aoki}
\affiliation{Institute for Materials Research, Tohoku University, Oarai, Ibaraki 311-1313, Japan}
\affiliation{Universit\'e Grenoble Alpes, CEA, IRIG, PHELIQS, F-38000 Grenoble, France}
\author{K.~Hiraki}
\affiliation{Department of Physics, Fukushima Medical University, Fukushima 960-1295, Japan}
\author{M.~Kimata}
\author{T.~Sasaki}
\affiliation{Institute for Materials Research, Tohoku University, Sendai 980-8577, Japan}
\date{\today}
\begin{abstract}
UTe$_2$ is a recently discovered spin-triplet superconductor.
One of the characteristic features of UTe$_2$ is a magnetic field ($H$)-boosted superconductivity above 16 T when $H$ is applied exactly parallel to the $b$ axis.
To date, this superconducting (SC) state has not been thoroughly investigated, and the SC properties as well as the spin state of this high-$H$ SC (HHSC) phase are not well understood.
In this study, we performed AC magnetic susceptibility and nuclear magnetic resonance (NMR) measurements and found that, up to 24.8 T, the HHSC state is bulk properties of \ce{UTe2} and quite sensitive to the $H$ angle, and that its SC character is different from that in the low-$H$ SC (LHSC) state.
The dominant spin component of the spin-triplet pair is along the $a$ axis in the LHSC state but is changed in the HHSC state along the $b$ axis.
Our results indicate that $H$-induced multiple SC states originate from the remaining spin degrees of freedom.   
\end{abstract}
\maketitle
Superconductivity occurs when a coherent quantum fluid is formed from electron pairs.
For most superconductors, although the total spin ($S$) of the pairs is in the singlet state ($S$ = 0), it is also possible in the triplet state ($S$ = 1).
Such superconductors, called ``spin-triplet superconductors,” are coherent quantum fluids with spin and orbital degrees of freedom.
Spin-triplet superconductors would involve rich physics, but are very rare.
Therefore, the nature of the spin-triplet pairing state was initially studied by analyzing the superfluidity of $^3$He \cite{OsheroffPRL1972,LeggettRMP1975}.
The recent discovery of ferromagnetic (FM) superconductors\cite{SaxenaNature2000, AokiNature2001, HuyPRL2007}, in which the ferromagnetism and superconductivity arise from same electrons, has made it possible to study the spin-triplet pairing state in the superconductors.
Additionally, a spin-triplet SC candidate UTe$_2$ has been newly discovered\cite{RanScience2019}; the SC transition temperature $T_{\rm c}$ is $1.6 \sim 2.0$ K\cite{Aoki2022review, Rosa2022}.
Although UTe$_2$ undergoes no FM transition, it was considered to be an end member of FM superconductors owing to its physical similarity to FM superconductors\cite{RanScience2019, Aoki2022review}. 
However, recent experimental results unveiled the presence of the incommensurate antiferromagnetic fluctuations as well as the FM fluctuations\cite{DuanPRL2020, KnafoPRB2021}.

The results of the nuclear magnetic resonance (NMR) Knight-shift ($K$) measurements to superconductors have provided important information about the spin state in the SC state\cite{YosidaPhysRev1957, MacLaughlin1976}.
However, in FM superconductors, such information is obscured because of the internal field produced by FM ordered moments.
Thus, UTe$_2$ provides a special opportunity for studying spin-triplet physics because the lack of FM moments means that precise $K$ measurements can be obtained.

\ce{UTe2} crystallizes \textit{Immm} space group ($\#71, D_{\rm 2h}$).
The possible SC symmetry and irreducible representation of spin-triplet superconductivity in a $D_{\rm 2h}$ point group corresponding to the orthorhombic crystal structure of UTe$_2$ in the zero field and $H \parallel b$ are listed in Table~\ref{t1} and Table~\ref{t2} \cite{IshizukaPRL2019,Aoki2022review}.

\begin{table}[htb]
\begin{center}
\caption{\label{t1}Classification of the odd-parity SC order parameters for point groups with $D_{\rm 2h}$ in a zero field. The irreducible representation (IR) and its basis functions are listed. The dominant spin component in the SC state is also shown.}
\vspace{3mm}
  \begin{tabular}{cccc}\hline \hline 
 \multicolumn{1}{c}{$D_{\rm 2h}$ (zero field)} \\ \hline 
IR & Basis functions & & SC spin comp.\\ \hline
$A_{\rm u}$ & $k_a \hat{a}$, $k_b \hat{b}$, $k_c \hat{c}$ & & \\
$B_{\rm 1u}$ & $k_b \hat{a}$, $k_a \hat{b}$ & & $c$ \\
$B_{\rm 2u}$ & $k_a \hat{c}$, $k_c \hat{a}$ & & $b$ \\
$B_{\rm 3u}$ & $k_c \hat{b}$, $k_b \hat{c}$ & & $a$ \\ \hline \hline
  \end{tabular}
  \end{center}
\end{table}
\begin{table}[htb]
\begin{center}
\caption{\label{t2}Classification of odd-parity SC phases occurring in UTe$_2$ under a $b$-axis magnetic field. The typical order parameters belonging to each IR are listed in Table \ref{f1}.}
\vspace{3mm}
  \begin{tabular}{cccc}\hline \hline 
 \multicolumn{1}{c}{IR of $C_{\rm 2h}$ (under field)} \\ \hline 
$H$ direction         & $A_{\rm u}^{H \parallel b}$                & &$B_{\rm u}^{H \parallel b}$               \\ \hline
$H \parallel b$ & $A_{\rm u}$ +$iB_{\rm 2u}$ & &$B_{\rm 3u} + iB_{\rm 1u}$ \\ \hline \hline
  \end{tabular}
  \end{center}
\end{table}

As a result of performing the NMR Knight shift measurements under low external fields\cite{NakamineJPSJ2019, NakaminePRB2021, NakamineJPSJ2021, FujibayashiJPSJ2022}, we found that UTe$_2$ is a spin-triplet superconductor with spin degrees of freedom.  
The important aspect to be clarified is the behavior of the remaining spin degrees under various experimental conditions, such as the application of a magnetic field and/or pressure.

The upper critical field of superconductivity ($H_{\rm c2}$) is strongly directionally dependent\cite{KnebelJPSJ2019, RanNatPhy2019}. 
When $H$ was perfectly aligned along the $b$ axis, $H$-boosted superconductivity was observed up to $\sim 35$ T \cite{RanScience2019, RanNatPhy2019, MiyakeJPSJ2019, KnebelJPSJ2019, KnafoCommPhys2021}. 
The use of microscopic measurements to investigate this high-$H$ SC (HHSC) state is critical for understanding the nature of spin-triplet superconductivity, as well as the SC mechanism of UTe$_2$.

A $^{125}$Te-enriched single crystal $5 \times 3 \times 1$ mm$^3$ in size, and with $T_{\rm c}$ $\sim 1.67$ K, was prepared by applying a chemical vapor transport method\cite{Aoki2022review}.
{The characterization of the present sample is described in Supplemental Materials (SM)\cite{supple}.
Figure \ref{f1}(a) shows the $^{125}$Te-NMR spectra for $H \parallel b$, which are plotted against $K = (f - f_0)/f_0 $. 
Here, $f$ is the NMR frequency and $f_0$ is the reference frequency determined as $f_0 = (^{125}\gamma_n/2\pi) \mu_0 H$ with a $^{125}$Te-nuclear gyromagnetic ratio $^{125}\gamma_n/2\pi = 13.454$~MHz/T. 
As shown in Fig.~\ref{f1}(b), there are two crystallographically inequivalent Te sites, $4j$ and $4h$, with the point symmetries $mm2$ and $m2m$ in \ce{UTe2}; these point symmetries are denoted as Te1 and Te2 sites, respectively. 
Correspondingly, we observed two $^{125}$Te NMR peaks, as has been reported previously \cite{TokunagaJPSJ2019}.
An NMR peak with a smaller [larger] $K$ in $H \parallel b$ was assigned as a Te(1) [Te(2)] peak, in accordance with a previous study \cite{TokunagaJPSJ2019,NakaminePRB2021}.

For the accurate alignment of the sample, we utilized the Te(1) NMR shift as an angle marker and an NMR probe with a two-axis rotator.
The two angles $\theta$ and $\phi$ have been defined as shown in Fig.~\ref{f1}(c); the sample orientation was adjusted by tuning $\theta$ and $\phi$ such that the Te(1) shift became the minimum value, as shown in Fig.~\ref{f1}(d) and (e).
The accuracy of the alignment was estimated to be $\pm 0.2^\circ$ for $\theta$ and $\pm 0.5^\circ$ for $\phi$, where $\theta$ ($\phi$) is the angle between the $b$ and $a$ ($c$) axes.
The details of how to align the samples is described in SM \cite{supple}.

\begin{figure}[tbp]
\begin{center}
\includegraphics[width=85mm]{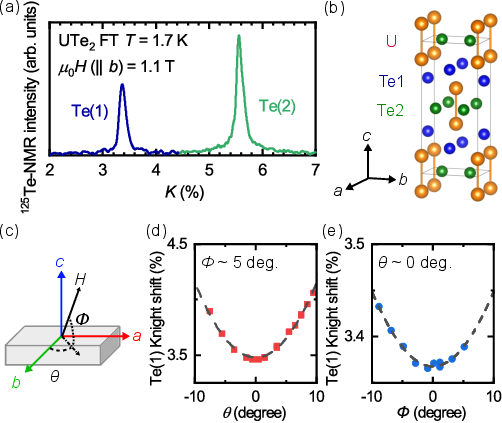}
\end{center}
\caption{(a) $^{125}$Te-NMR spectra measured in $H \parallel b$. A NMR peak with the smaller [larger] $K$ is called Te(1) [Te(2)] in this paper. (b) The crystal structure of UTe$_2$\cite{VESTA}. There are two Te sites in UTe$_2$. (c) Definition of the angles $\theta$ and $\phi$ against the crystalline axes in UTe$_2$ (d), [(e)] The $\theta$ [$\phi$] dependence of the $K$ in the Te(1) peak. The $b$ axis [($\theta$, $\phi$) = (0, 0)] is determined from the minimum of $K$ at the Te(1) site.}
\label{f1}
\end{figure}

To confirm the SC phase diagram, we measured the tuning frequency ($\nu_{\rm tune}$) and radio frequency (RF) reflection coefficient for the NMR tank $LC$ circuit using a vector network analyzer, where $L$ and $C$ are the inductance and  capacitance, respectively.
$\nu_{\rm tune} \sim 1/\sqrt{LC}$ is a good measure for tracking the superconductivity, because the inductance of the NMR coil with the sample, i.e., $L = L_{0} (1 + q \chi_{\rm AC})$, where $q$ is the ``filling factor,'' changes at the SC onset.
Thus, the change in AC susceptibility ($\chi_{\rm AC}$) due to SC diamagnetism can be detected {\it in situ} by measuring the change in $\nu_{\rm tune}$ across $T_{\rm c}$ or $H_{\rm c2}$.

Figure \ref{f2}(a) shows the variation in $-\Delta \nu/\nu_{\rm tune}$, as measured by sweeping $H$ at 1.5, 1.0, and 0.6 K.
At 1.5 and 1.0 K, the SC transitions were indicated by sudden decreases in the fields, as shown by the arrows.
Although UTe$_2$ is in the SC state at 0.6 K, $-\Delta \nu/\nu_{\rm tune}$ was found to exhibit a characteristic $H$ dependence.  
Increasing $H$ above 14 T corresponded to sharp decreases in $|-\Delta \nu/\nu_{\rm tune}|$; however, further increase beyond 16.5 T coincided with increasing $|-\Delta \nu/\nu_{\rm tune}|$, indicating a kink at $H_{\rm kink} \sim 16.5$ T.
Figure \ref{f2}(b) shows the $T$ dependence of $-\Delta \nu/\nu_{\rm tune}$ for $\mu_0 H$ = 7.5, 15.5, 16.5, and 24 T on cooling.

The minimum value of $|-\Delta \nu/\nu_{\rm tune}|$, in relation to SC diamagnetism, was observed at 16~T, consequently demonstrating the same tendency as $T_{\rm c}$.
This indicates that the HHSC state is bulk properties of \ce{UTe2}.
The $H$ and $T$ dependencies of $-\Delta \nu/\nu_{\rm tune}$ suggest that the SC character changed at $\mu_0 H_{\rm kink} \sim 16.5$ T.
In fact, the HHSC state was found to be very sensitive to the angle $\theta$. Additionally, superconductivity was observed within $\pm 3^{\circ}$ at 24 T, as shown in Fig.~\ref{f2}(c), and the unexpected minute $\theta$ rotation ($\Delta\theta \sim 4 ^{\circ}$) that occurred during the experiments completely suppressed the high-$H$ superconductivity, although the SC diamagnetism in the low-$H$ SC (LHSC) state was nearly unchanged, as shown by the dotted curve in Fig.~\ref{f2}(a).     
This result is in good agreement with the results presented in previous reports\cite{KnebelJPSJ2019, RanNatPhy2019}.
Based on the $T$- and $H$-scan measurements of $-\Delta \nu/\nu_{\rm tune}$, we developed the SC $H_{\rm c2}$ phase diagram shown in Fig.~\ref{f2}(d).
The $H_{\rm kink}$ anomaly, like the overall phase diagram, is consistent with the phase transitions determined by the recent specific-heat measurements\cite{RosuelarXiv2022}.
Because the responses to $H$ and $\theta$ are different between the LHSC and HHSC states, it is reasonable to consider that the kink at $H_{\rm kink}$ marks a phase transition between the two SC states.   
Such a transition between two bulk SC states was also confirmed in the work\cite{RosuelarXiv2022} by linear magnetostriction and thermal dilatation, evidencing anomalies due to vortex pinning in both phases.
The results of the $H$-sweep measurement at 0.6 K revealed clear hysteresis behavior at $\mu_0 H^{*} \sim 4$ T; it was found to be related to an anomaly of the vortex state, because the anomaly was not previously observed in the $H$ dependence results for the electronic term in specific-heat measurements\cite{KittakaPRR2020}. 
The details of this anomaly have been studied and will be reported in a separate paper.

\begin{figure}[tbp]
\begin{center}
\includegraphics[width=85mm]{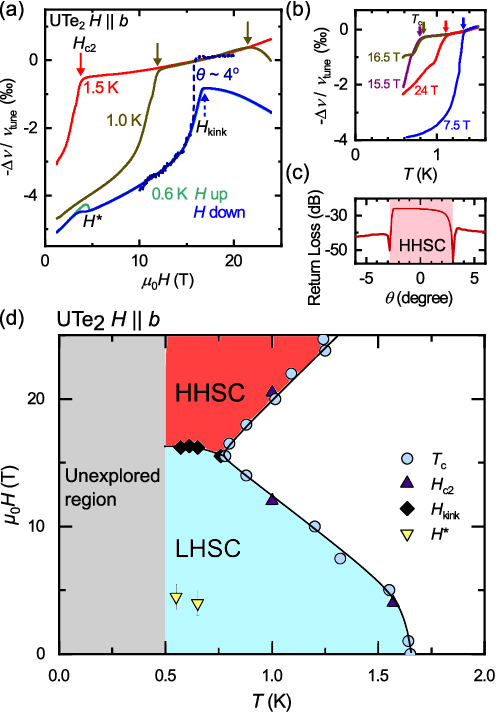}
\end{center}
\caption{(a) $H$ dependence of $-\Delta \nu/\nu_{\rm tune}$ values up to 24.8 T, as measured at 0.6, 1.0 and 1.5 K. At 0.6 K, $-\Delta \nu/\nu_{\rm tune}$ exhibited a kink at $H_{\rm kink}$; the $H$ dependence of $-\Delta \nu/\nu_{\rm tune}$, which was determined by performing $H$-up and $H$-down sweeps, is shown. The dotted curve shows the $H$ dependence of $-\Delta \nu/\nu_{\rm tune}$ when the minute $\theta$ rotation unexpectedly occurred in the sample. (b) Temperature dependence of $- \Delta \nu/\nu_{\rm tune}$ in relation to the AC magnetic susceptibility $\chi_{\rm AC}$, as measured at 7.5, 15.5, 16.5, and 24 T. $T_{\rm c}$ in the field is denoted by the arrow with the same color. (c) Angle dependence of the return loss of the NMR tank circuit at 24 T. When the sample was in the SC state, the quality factor of the circuit $Q$ was lower owing to the change in the impedance of the circuit. High-$H$ superconductivity was observable within $\pm 3 ^{\circ}$. (d) SC upper critical field $H_{\rm c2}$ determined by performing $T$- and $H$-scan measurements of $- \Delta \nu/\nu_{\rm tune}$.}
\label{f2}
\end{figure}

To investigate the SC properties, particularly the spin susceptibility in the HHSC state, we performed $^{125}$Te NMR measurements at the Te(2) peak with a larger $K$.
Figures \ref{f3}(a) and (b) show the Te(2) NMR spectra measured at various temperatures below 2.5 K at 1 and 24 T, respectively.
At 1 T, the single-peak spectrum gradually shifted to the low-$K$ side in the normal state and sharply shifted immediately below $T_{\rm c}$; this was accompanied by spectrum broadening.
The $^{125}$Te NMR spectrum measured under conditions of 24 T and 2.5 K revealed a double-peak structure that is attributable to its high resolution; the right peak was found to have a 0.04 \% larger $K$ than the main peak.
The $H$ dependence of the Te(2)-NMR spectrum is shown in SM\cite{supple}.
Several possibilities were considered for the origin of the larger-$K$ peak; they include the occurrence of a mosaic structure and/or minute U-atom deficiency in the single-crystal sample. 
In the former case, the misalignment of the mosaic was estimated to be 2.0$^{\circ}$ (8.1$^{\circ}$) on the $a$ ($c$) axis; additionally, the $^{125}$Te NMR measurement for the higher-$T_{\rm c}$ single crystal is critical for the latter possibility because $T_{\rm c}$ seems to be very sensitive to a U-atom deficiency\cite{Aoki2022review,Rosa2022,Haga_2022}.
Further experiments are required to clarify the origin of the larger $K$ peak.
As $T$ was decreased, the two peaks gradually shifted to the lower-$K$ side in the same manner.
Because the resolution of the higher-$K$ peak is not sufficient for analysis, we focus on the main peak shown by the arrows.

\begin{figure*}[t]
\begin{center}
\includegraphics[width=\linewidth]{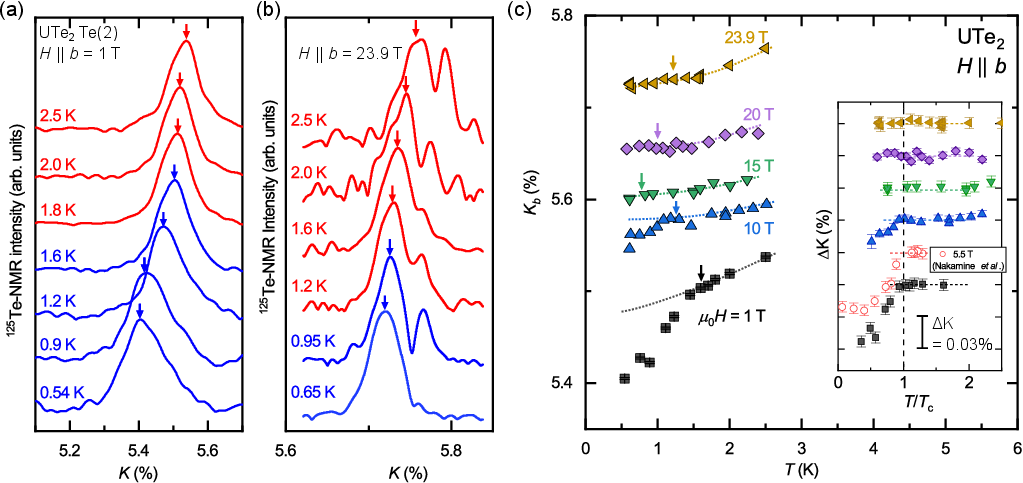}
\end{center}
\caption{Te(2) NMR peaks measured at various temperatures below 2.5 K at $\sim 1$ T (a) and $\sim 24$ T (b). (c) Temperature dependence of $K$, as determined by analyzing the Te(2) NMR peak shown by the arrow. 
The dotted line is the normal-state behavior extrapolated using the 2nd-order polynominal function as explained in the supplemental materials.
(inset) Temperature dependence of the change in the Knight shift from the normal state. The temperature dependence of the normal-state Knight shift was fitted and subtracted using a quadratic function. The horizontal dashed line represents the baseline, and the vertical dashed line represents the transition temperature. }
\label{f3}
\end{figure*}
Figure \ref{f3}(c) shows the temperature dependence of $K$ of the main peak, as determined from the NMR spectra measured at 1, 10, 15, 20, and 24 T.
A decrease in $K$ was clearly observed at $T_{\rm c}$ for 1 and 10 T; the magnitude and $H$ dependence of the $K$ decrease below $T_{\rm c}$ are in agreement with previous results\cite{NakaminePRB2021, NakamineJPSJ2021}.
In contrast, when 20 and 24 T were applied, $K$ gradually decreased at temperatures below 2.5 K without any appreciable anomaly at $T_{\rm c}$($H$).
To quantify the Knight-shift decrease ($\Delta K$) ascribed to the superconductivity, the normal-state $T$ dependence was subtracted from the observed $K_b$ and $\Delta K$ was plotted for each $H$, as shown in the inset of Fig.~\ref{f3}(c).
$\Delta K$ was near zero in the SC state for values above 15~T, although this field is still in the LHSC.
This behavior is consistent with previous NMR measurements\cite{NakamineJPSJ2021}.
A similar $\Delta K \sim 0$ trend was previously observed in the $a$-axis Knight-shift measurement results for the LHSC state, where the dominant SC spin component occurred along the $a$ axis \cite{FujibayashiJPSJ2022}.
Thus, these results indicate that $b$-axis spin-polarized superconductivity is induced by a $b$-axis magnetic field.

We will now discuss possible SC states in the HHSC region.
Considering the observed spin-susceptibility and field-boosted behavior, the ground state of the HHSC is $A_{\rm u}^{H \parallel b}$, as presented in table~\ref{t2}; this is because the SC spin component is parallel to $H \parallel b$ in the HHSC region.
This is consistent with the theoretical suggestion\cite{IshizukaPRL2019,ShishidouPRB2021}.
Although the $\Delta K$ changes smoothly, the kink anomaly in the field dependence $-\Delta \nu/\nu_{\rm tune}$ implies a phase transition between the HHSC and LHSC states; thus, the LHSC state is determined to be $B_{\rm u}^{H \parallel b}$.
These results strongly support the $B_{\rm 3u}$ scenario at a low-field limit\cite{NakamineJPSJ2019,NakaminePRB2021,NakamineJPSJ2021,FujibayashiJPSJ2022}.
Under $H \parallel b$, $B_{\rm 3u}$ at zero field becomes $B_{\rm u}^{H \parallel b}$ with crossover  (without any transition).
Note that, for thermodynamic limitation, it is not allowed the ``tri-critical point'' with three 2nd order phase transition line.
Thus, the phase transition line inside the SC region should be 1st order or there is a hidden phase transition line with 2nd order phase transition \cite{IshizukaPRL2019}.
The results of the crude up-down $H$-sweep measurement of $- \Delta \nu/\nu_{\rm tune}$ at 0.6 K revealed the occurrence of one kink without any hysteresis near $H_{\rm kink}$[Fig.~\ref{f2}(b)]; this seems to exclude the ``1st order'' phase transition scenario.
Rather we suggest the presence of another phase transition line characterized with the SC properties of the HHSC state such as $\Delta K = 0$.
Further precise NMR measurements are required to understand the relationship between the LHSC and HHSC pahses.

In addition, it is noteworthy that the enhancement of the superconductivity against $H$ was found to be stronger than that previously reported \cite{KnebelJPSJ2019, RanScience2019, KnafoCommPhys2021}.
Because the value of $T_{\rm c}$ at $H = 0$ for the current sample (1.67 K) was slightly higher than that of previous samples ($\sim$ 1.5 K), the upturn behavior is seemingly dependent on the sample quality, suggesting the intrinsic properties of UTe$_2$.   
A similar level of superconductivity robustness by $H \parallel b$ was observed in the FM superconductors URhGe\cite{LevyScience2005} and UCoGe\cite{AokiJPSJ2009}, in which critical FM fluctuations were determined to play an important role \cite{TokunagaPRL2015, HattoriJPSJ2014,Wu2017,IshidaPRB2021}. 
Because superconductivity occurs in the paramagnetic state of UTe$_2$, such critical FM fluctuations were not anticipated.
Alternatively, we speculate that 
the critical fluctuations related to the incommensurate antiferromagnetic fluctuations\cite{DuanPRL2020, KnafoPRB2021}, which may be induced by $H \parallel b$ above $16.5$~T, plays an important role in the mechanism governing the HHSC state. 
It is interesting that the SC pairing interaction can be tuned by adjusting the $H$ applied along the $b$ axis; this seems to be a common feature of U-based FM and nearly FM superconductors with Ising anisotropy under normal-state magnetic conditions, although the SC pairing interaction is not clarified in UTe$_2$.   

In conclusion, we have determined from the results of $in$-$situ$ $\chi_{\rm AC}$ and NMR measurements at magnetic field strengths up to 24.8 T, that the HHSC state is bulk properties of \ce{UTe2}, and that the spin component of the triplet pair occurs along the $b$ axis in the HHSC state, which is different from that in the LHSC state.
The results presented here provide decisive evidence that the spin degrees in a spin-triplet pair can be controlled by an external magnetic field $H$.
This is a unique phenomenon that is not expected in spin-singlet superconductors, but is inherent to spin-triplet superconductors.
Exploring unique phenomena related to the spin degrees of freedom in spin-triplet superconductors is important because this information can facilitate their application.
This study is currently in progress.

The authors thank Y. Yanase, K. Machida, S. Fujimoto, V. P. Mineev, Y. Maeno, S. Yonezawa, J-P. Brison, G., Knebel, and J. Flouquet for their valuable discussions.
They would also like to thank K. Shirasaki and M. Nagai for their technical support and Editage (www.editage.com) for English-language editing. 
This work was performed at High Field Laboratory for Superconducting Materials (HFLSM) under the GIMRT program of the Institute for Materials Research (IMR), Tohoku University (Proposal Nos. 202012-HMKPB-0012, 202112-IRKAC-0023, and 202112-HMKPB-0008). 
This work was supported by the Kyoto University LTM Center and Grants-in-Aid for Scientific Research (Grant Nos.JP19K03726, JP19K14657, JP19H04696, JP19H00646, JP20H00130, JP20KK0061, and 21K18600).
This work was also supported by JST SPRING (Grant Number JPMJSP2110).
Some authors would like to acknowledge the support from the Motizuki Fund of Yukawa Memorial Foundation.

\bibliographystyle{apsrev4-2}

\end{document}